\documentstyle[12pt,twoside,fleqn,espcrc1,epsfig]{article}

\newcommand{\AmS}{{\protect\the\textfont2
  A\kern-.1667em\lower.5ex\hbox{M}\kern-.125emS}}

\title{Search for Disoriented Chiral Condensates in 158$\cdot$A GeV
             Pb+Pb Collisions}

\author{\normalsize{Tapan K. Nayak$^c$
\footnote{email : nayak@veccal.ernet.in}}
\it for the WA98 Collaboration }

\begin{document}

\maketitle

\noindent
\small{M.M.~Aggarwal$^{a}$, A.~Agnihotri$^{b}$, Z.~Ahammed$^{c}$,
A.L.S.~Angelis$^{d}$, V.~Antonenko$^{e}$, 
V.~Arefiev$^{f}$, V.~Astakhov$^{f}$,
V.~Avdeitchikov$^{f}$, T.C.~Awes$^{g}$, P.V.K.S.~Baba$^{h}$
% ...
}

\smallskip
\noindent
\small\it{$^{a}$Univ. of Panjab (India)}
\small\it{$^{b}$Univ. of Rajasthan (India)}
\small\it{$^{c}$VECC, Calcutta (India)}
\small\it{$^{d}$Univ. of Geneva (Switzerland)}
\small\it{$^{e}$Kurchatov (Russia)}
\small\it{$^{f}$JINR, Dubna (Russia)}
\small\it{$^{g}$ORNL, Oak Ridge (USA)}
\small\it{$^{h}$Univ. of Jammu (India)}
\small\it{$^{i}$Univ. of M{\"u}nster (Germany)}
\small\it{$^{j}$SUBATECH, Nantes (France)} 
\small\it{$^{k}$GSI, Darmstadt (Germany)}
\small\it{$^{l}$NIKHEF, Utrecht (The Netherlands)}
\small\it{$^{m}$Univ. of Lund (Sweden)}
\small\it{$^{n}$Univ. of Tsukuba (Japan)}
\small\it{$^{o}$NPI, Rez (Czech Rep.)}
\small\it{$^{p}$KVI, Groningen (The Netherlands)}
\small\it{$^{q}$INS, Warsaw (Poland)}
\small\it{$^{r}$MIT, Cambridge (USA)}
\small\it{$^{s}$IOP, Bhubaneswar (India)}
\small\it{$^{t}$Univ. of Tennessee (USA)}

\normalsize



\bigskip

   
The formation of Disoriented Chiral Condensates (DCC) \cite{raj}
in high energy heavy-ion collisions is
associated with large event-by-event fluctuation in the 
ratio of neutral to charged pions.
The probability distribution of neutral pion fraction, $f$, in a 
DCC domain has been shown \cite{anselm} to follow the relation:
\begin{equation}
P(f) =  \frac{1}{2\sqrt{f}} {\rm ~~~~~~~~~~where~~~~} f = N_{\pi^0}/N_{\pi}
\end{equation}
which is quite different from that of the normal pion production mechanism.
The detection and study of DCC is expected to provide valuable 
information about the chiral phase transition and vacuum structure of strong 
interactions.

Theoretical predictions suggest that the isospin 
fluctuations caused by DCC would produce clusters
of coherent pions in phase space forming domains localized in
phase space. However, experimental studies \cite{WA98-3}
carried out so far have been restricted to global fluctuation
in the number of charged particles and photons.
Here we present first experimental results
to search for localized domains of DCC
based on event-by-event fluctuation in the relative 
number of charged particles and photons detected within the acceptance of
the WA98 experiment. 


    In the WA98 experiment, the main emphasis has 
    been on high precision, simultaneous detection of both hadrons 
    and photons. Charged particle hits ($N_{\rm ch}$) 
    were counted using a circular Silicon Pad 
    Multiplicity Detector (SPMD) \cite{WA98-3} located 32.8 cm downstream 
    of the target with a coverage of $2.35< \eta < 3.75$.
    Photon multiplicities ($N_{\gamma-like}$)
    were measured by 
    the preshower Photon Multiplicity Detector (PMD) \cite{WA98-9}
    placed at 21.5 meters downstream of the target covering and $\eta$
    range of $2.9-4.2$.
    It consisted of an array of plastic scintillator pads, arranged
    inside 28 box modules, placed behind 3$X_0$ thick lead converter 
    plates. 
    A total of 66K 
    central events corresponding to top 5\% of the minimum bias
    cross section, determined from the total transverse energy measured 
    by the midrapidity calorimeter (MIRAC), were analyzed. Common
    coverage of PMD and SPMD was considered.


    Physics analysis with the cleaned data was performed by
    comparing with mixed and simulated events. Mixed events were
    constructed from data combining hits from PMD and SPMD 
    chosen randomly from different events taking care of 
    the two track resolutions while maintaining the correlation between
    $N_{\gamma-like}$ and $N_{\rm ch}$, event-wise, as in the data.

    Simulated events were generated using VENUS 4.12 event 
    generator with default parameters. The output was then processed 
    through a detector simulation package in the GEANT 3.21
    framework. This incorporates the full WA98 experimental setup.
    The effect of Landau fluctuations in the
    energy loss of charged particles in silicon \cite{WA98-3} was
    included in the SPMD simulation.
    For PMD simulation the GEANT results in terms of energy deposition in 
    pads were converted to the pad ADC  values taking into
    account the detector and readout effects.
    At this stage the simulated data (henceforth referred to as VENUS) and
    experimental data were ready to be 
    processed with common analysis tools to study specific physics issues.


\medskip

\noindent
\underline{$N_\gamma$ and $N_{\rm ch}$ CORRELATION}

   The correlation between $N_{\gamma-like}$
   and $N_{\rm ch}$ has been studied 
   in smaller $\phi$-segments by dividing the
   $\phi$-space into 2, 4, 8 and 16 bins using the method
   described in Ref. \cite{WA98-3}.
   The correlation plots of $N_{\gamma-like}$ and $N_{\rm ch}$ have been
   constructed for each $\phi$ bin.
   A common correlation axis ($Z$) has been obtained by fitting
   the above distributions with a second order polynomial.
   The closest distance ($D_{Z}$) of the data points to the
   correlation axis has been calculated numerically with the
   convention that $D_{Z}$ is positive for points below the 
   Z-axis.
   The distribution of $D_Z$ represents the
   relative fluctuations of $N_{\gamma-like}$ and $N_{\rm ch}$ 
   from the correlation axis at any given $\phi$ bin.
   In order to compare these fluctuations at different scales in the
   same level, we work with a scaled variable,
   $S_{Z} = D_Z/\sigma(D_Z)$, where $\sigma(D_Z)$ corresponds
   to VENUS events.

\begin{figure}[t]
\setlength{\unitlength}{1mm}
\begin{picture}(140,56)
\put(0,0){
\epsfxsize=8.0cm
\epsfysize=7.5cm
\epsfbox{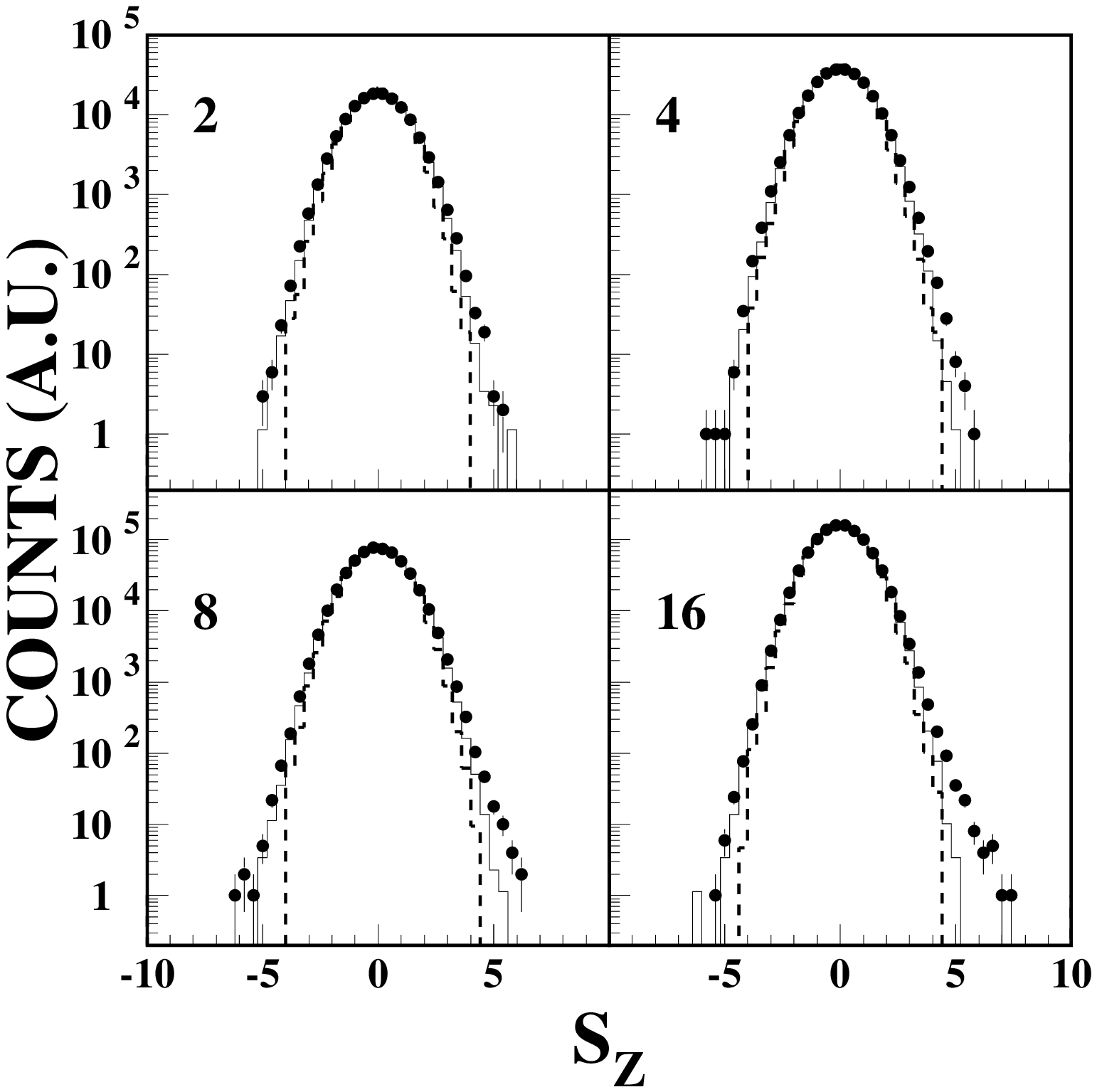}
}
\put(75,0){
\epsfxsize=7.5cm
\epsfysize=6.5cm
\epsfbox{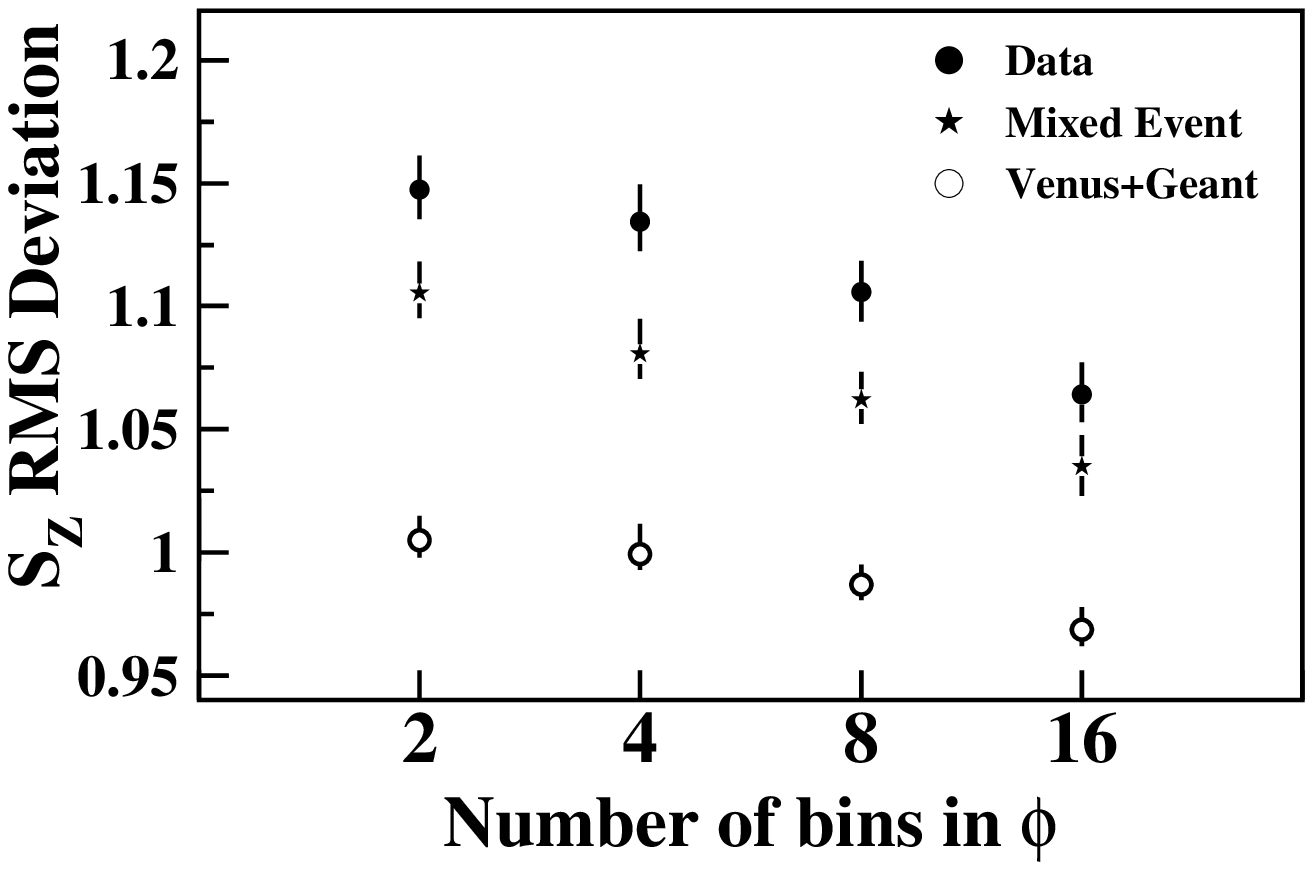}
}
\end{picture}
\vspace{-1.0cm}
\caption{
(a) 
$S_Z$ distributions for 2,4,8 and 16 divisions in $\phi$ angle.
Solid circles are for data, solid histograms are for mixed events
and dashed histograms are for simulation.
(b) 
Corresponding root mean square (rms) deviations. 
The bars represent systematic errors.
}
\end{figure}

   The $S_{Z}$ distributions  
   are shown in the left panel of Figure~1. The corresponding
   rms deviations for different $\phi$
   bins are shown in the right of panel of figure~1.
   The presence of events with DCC domains of a particular size would 
   show up as deviations from general correlation line at the proper 
   bin. This would result in a broader distribution of $S_Z$ compared 
   to those for generic events at that particular division.
   From figure~1(b) we also observe that the rms deviations of
   data differ significantly from those of mixed events 
   when the number of bins in $\phi$ is 8 or less.
   This indicates the presence of
   non-statistical fluctuations in $N_{\gamma-like}$ and $N_{\rm ch}$
   for bin sizes $\ge 45^0$.

\medskip

\noindent
\underline{DISCRETE WAVELET TECHNIQUE}

	The novel and powerful method of discrete wavelet 
	transformation (DWT) has also been used for the analysis of 
	the data. 
   Simulation studies by Huang et. al. \cite{huang} show that 
   the DWT analysis could be a powerful technique for the search of 
   localized DCC.
   For the present DWT analysis the azimuthal space of the PMD and the
   SPMD were divided into small bins in $\phi$, the 
   number of bins in a given scale $j$ being $2^j$.
   The input to the DWT analysis is a spectrum of the
   sample function at the smallest bin in
   $\phi$ corresponding to the highest resolution scale, $j_{max}=5$.
   Our sample function was chosen to be of the form
\begin{equation}
    f^\prime = \frac {N_{\gamma-like}}{N_{\gamma-like}+N_{\rm ch}}
\end{equation}
   The output of the DWT consists of a set of wavelet or father function 
   coefficients (FFCs) at each scale, from $j=1$,...,($j_{max}-1)$. 

   The distribution of FFCs for a generic distribution of  particles 
   is gaussian in nature \cite{huang,dccflow,dccstr}. However, the
   presence of DCC-like fluctuation makes the distribution
   non-gaussian, with a larger rms deviation of the FFC distribution.
   Comparing the rms deviations of the FFC distribution
   of data, mixed and simulated events one can get an idea about the 
   localized
   fluctuations in the distributions of 
   $N_{\gamma-like}$ and $N_{\rm ch}$ in the azimuthal space.

\begin{figure}[t]
\setlength{\unitlength}{1mm}
\begin{picture}(140,56)
\put(0,0){
\epsfxsize=8.0cm
\epsfysize=7.5cm
\epsfbox{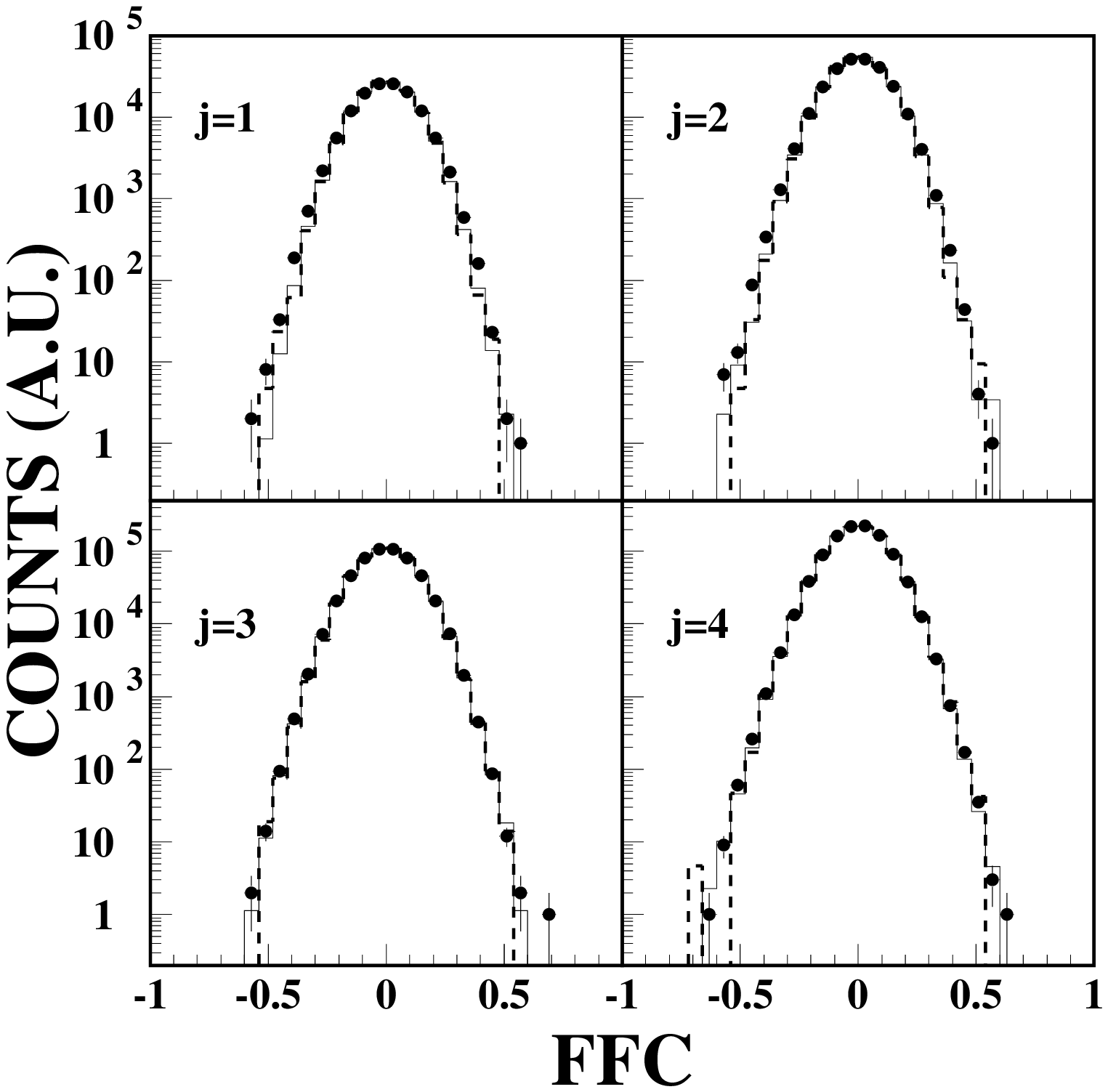}
}
\put(75,0){
\epsfxsize=7.5cm
\epsfysize=6.5cm
\epsfbox{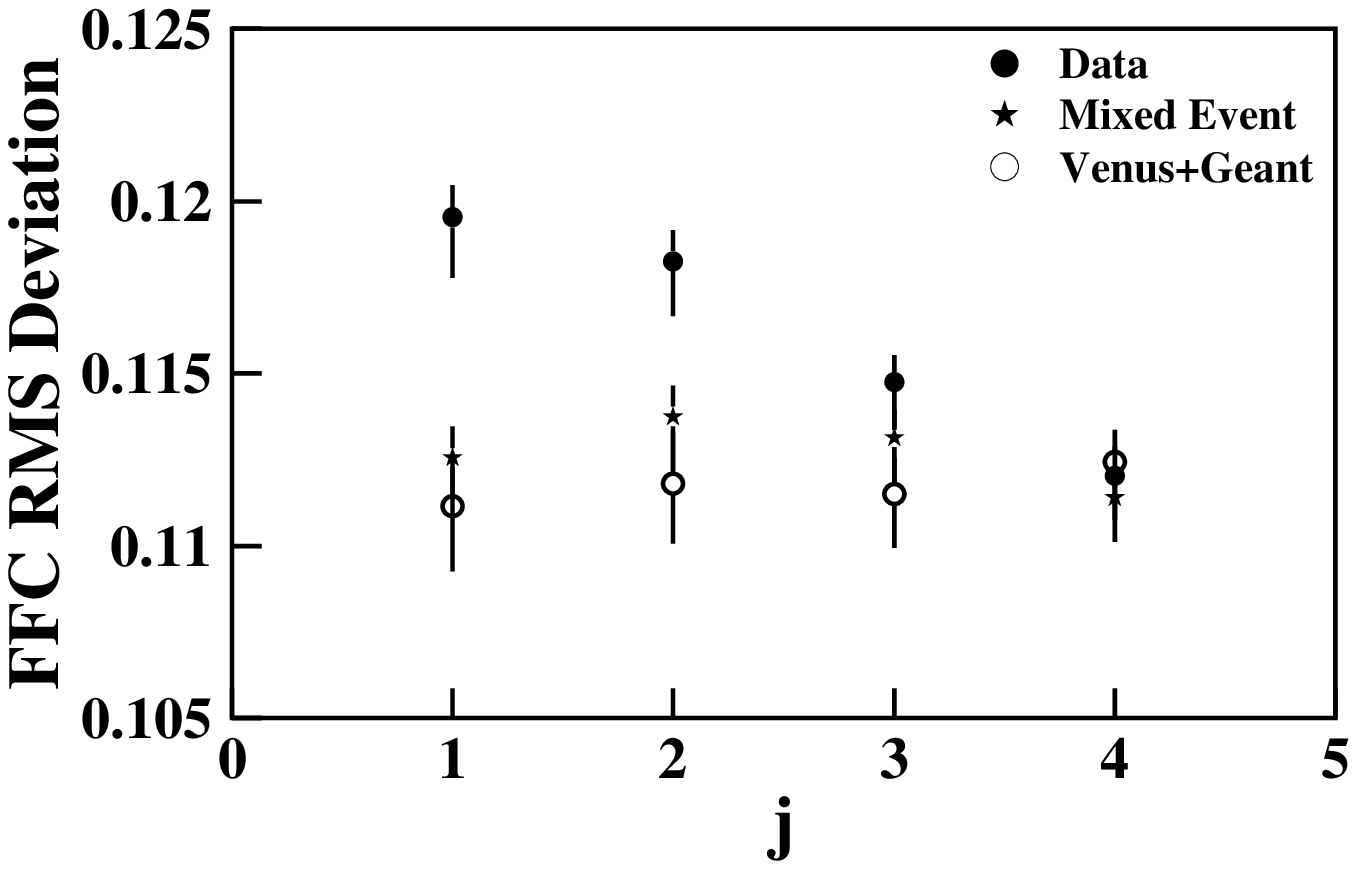}
}
\end{picture}
\vspace{-1.0cm}
\caption{
(a) 
FFC distributions at $j=1-4$.
Solid circles are for data, solid histograms are for mixed events
and dashed histograms are for simulation.
(b) 
Corresponding root mean square (rms) deviations at scales
$j=$ 1-4. The bars represent systematic errors.
}
\end{figure}

   Figure~2(a) shows the FFC
   distributions at scales $j=$1-4 for data, mixed events and VENUS.
   The corresponding rms deviations, $\xi$,
   are shown in figure~2(b). 
   We observe that
   the rms deviations of data, mixed and VENUS
   events match with each other at scales $j=$ 3 and 4.
   For scales $j=1$ and 2, the rms deviations of VENUS and mixed
   events are close to each other whereas there is a clear
   difference of these values from those of the data.
   In the DWT analysis, the information about the degree of fluctuation at a 
   scale $j$ is obtained from FFC distribution
   at scale $j+1$. One can thus 
   deduce that the differences seen at $j=2$ and 1
   are due to the fluctuations in $N_{\gamma-like}$ and $N_{\rm ch}$
   at scale $j=3$, i.e. around 45$^\circ$ in $\Delta\phi$.


    A detailed analysis of the $\eta-\phi$ phase
    space distribution of charged particles and photons have been
    carried out using two independent methods.
    Both of these indicate the presence of nonstatistical fluctuations
    in localized regions in terms of photon and charged particle 
   multiplicities. This could have contributions from many different
   physics effects. Attempts are being made
   to explain this
   assuming the fluctuations to be solely due to the presence
   of localized domains of DCC.

\end{document}